# Emerging Mobile Phone-based Social Engineering Cyberattacks in the Zambian ICT Sector

**Aaron Zimba[1], George Mukupa[2], Victoria Chama[3]**
[1]Department of Computer Science and Information Technology, Mulungushi University
[2]Department of Science and Mathematics, Mulungushi University
Emails; azimba@mu.ac.zm, georgemukupa@mu.ac.zm, vchama@mu.ac.zm

Abstract: The number of registered SIM cards and active mobile phone subscribers in Zambia in 2020 surpassed the population of the country. This clearly shows that mobile phones in Zambia have become part of everyday life easing not only the way people communicate but also the way people perform financial transactions owing to the integration of mobile phone systems with financial payment systems. This development has not come without a cost. Cyberattackers, using various social engineering techniques have jumped onto the bandwagon to defraud unsuspecting users. Considering the aforesaid, this paper presents a high-order analytical approach towards mobile phone-based social engineering cyberattacks (phishing, SMishing, and Vishing) in Zambia which seek to defraud benign victims. This paper presents a baseline study to reiterate the problem at hand. Furthermore, we devise an attack model and an evaluation framework and ascertain the most prevalent types of attack. We also present a logistic regression analysis in the results section to conclude the most prevalent mobile phone-based type of social engineering attack. Based on the artifacts and observed insights, we suggest recommendations to mitigate these emergent social engineering cyberattacks.

## 1. Introduction

Information and Communication Technologies (ICTs) have touched almost every area of our daily lives from the way we communicate to the way we acquire and use different goods and services. It is indisputable that this paradigm shift has come with tremendous benefits (Tudorel and Vintila 2020). With advances in technologies and the advent of the Fourth Industrial Revolution (Xu, David, and Kim 2018), mobile phones have gone beyond their traditional role as mere communication tools but are now used for critical tasks such as financial transactions not limited to mobile money and Internet banking (Alhassan et al. 2020). This is evidenced by the fact that they are one of the few innovative ICT technologies that have found their way amongst the poor (Gebremariam 2020).

In Zambia, the liberalisation of the telecoms sector and the subsequent introduction of multiple mobile network providers saw the widespread adoption of mobile phones (Zimba et al. 2020). Within the past decade, as shown in Figure 1, the number of active mobile phone subscribers has grown from half the population to more than 100%. As of the first quarter (Q1) of 2021 according to ZICTA statistics (ZICTA 2021), the number of active mobile phone subscribers stood at 19.2 million against a population of about 18.4 million and this represents a penetration rate of about 104.7%. Additionally, the number of mobile Internet users is about 10.4 million indicating that 56.3% of the population actively uses Internet via mobile phones. Nonetheless, this tremendous adoption of mobile phones and Internet usage hasn't come without a cost.

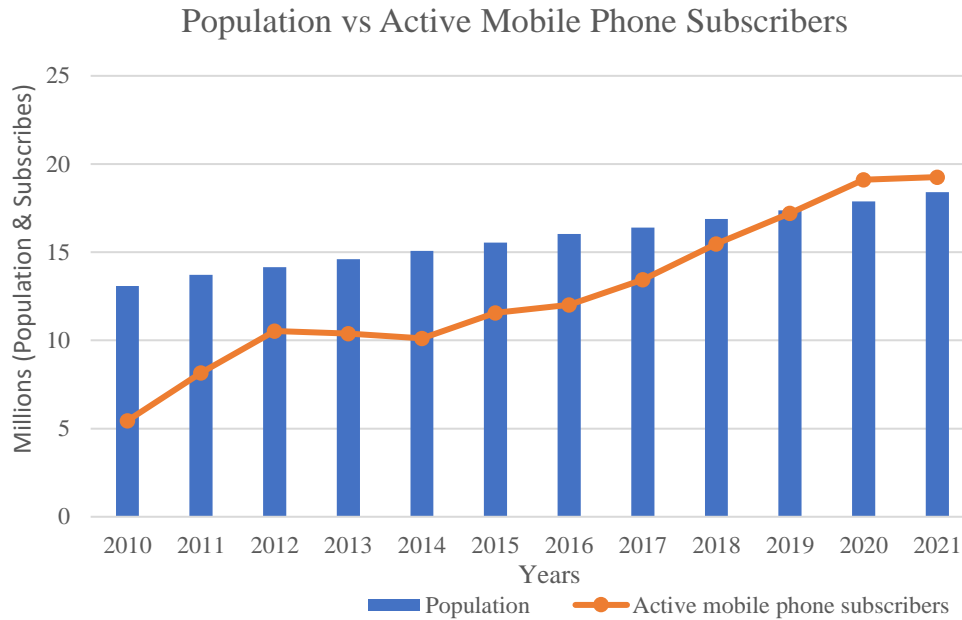

Figure 1. Population growth vs number of active mobile subscribers in the last decade (ZICTA 2021)

As is the case of every environment in which a disruptive technology thrives, the adverse effects of the widespread adoption and usage of mobile phones are beginning to emerge (Apau, Koranteng, and Gyamfi 2019). One prominent problem that has arisen is cyberattacks, which have led to the subsequent enactment of the Cybersecurity Bill of 2021 (GRZ 2021).

Cyberattacks come in different forms not limited to password attacks, Denial of Service (DoS), Man-In-The-Middle (MITM), social engineering, etc. Social engineering stands out in the mobile phone landscape because it requires little technical knowledge or tools and is feasible both on feature and smartphones (Salahdine and Kaabouch 2019). In this type of attack, the attacker aka social engineer uses psychology to manipulate the benign potential victim by leveraging some form of social influence (Bullée et al. 2018). A particular type of social engineering attack which is the focus of this paper is phishing. Phishing is a subcategory of social engineering attacks usually used to steal a victim's sensitive information such as login credentials, mobile money PIN codes, credit card information, etc (Shirazi et al. 2021). This is usually effective when the attacker impersonates or masquerades as a trusted entity, then dupes the victim into doing what an SMS says, clicking on a link in WhatsApp, following the instructions of the masquerading attacker over a voice call, etc. The victim is misled into performing an action that will result in a financial loss or breach of sensitive information such as passwords or PINs.

Since the attacker uses the phone as an attack medium, he has to select a way of interacting with the potential victim, i.e., either by text or by voice. In the case of the former, it can be a misleading SMS to trick the potential victim to authorise or perform a malicious action that would lead to loss of money or leaking of private information. Since this type of phishing leverages an SMS as an attack vector, it is called SMishing (Mishra and Soni 2019). In the case of the latter, the attacker makes a phone call to a would-be victim and impersonates an authority such as a bank or a mobile phone call centre operator. Since this type of phishing leverages a voice call, it is called Vishing (Jones et al. 2020).

In Zambia, the rapid growth of cybercrime is attributed to the widespread adoption of mobile phones which has exposed every user to cyberattacks (Xinhua News 2020). Mobile phones enable users to perform mobile money transactions or electronic payments which is a much safer way of transacting especially in this era of Covid19. Mobile money usage in Zambia has continued to grow tremendously with the value of transactions more than doubling in 2019 at ZMW 49.6 billion from ZMW 22.2 billion in 2018, whereas the volume of transactions grew to 553 million from 304 million for the same period. With the worsening of the Covid19 pandemic which has seen more users adopt mobile transactions, this trend has continued to repeat itself echoing a 750.5 million volume of mobile money transactions in 2020 with the corresponding value of ZMW 105.8 billion (BoZ 2021). These volumes, both in the number of transactions and value, surpass conventional bank transactions via Electronic Funds Transfer (EFT)/Direct Debit and Credit Clearing (DDACC) by huge margins which stood at 8.1 million and ZMW 78 billion in 2021 (BoZ 2021) respectively. The diagram in Figure 2 shows how mobile money has in the last decade come to take the lead as the major type of financial transaction in Zambia against Electronic Funds Transfer (EFT)/Direct Debit and Credit Clearing.

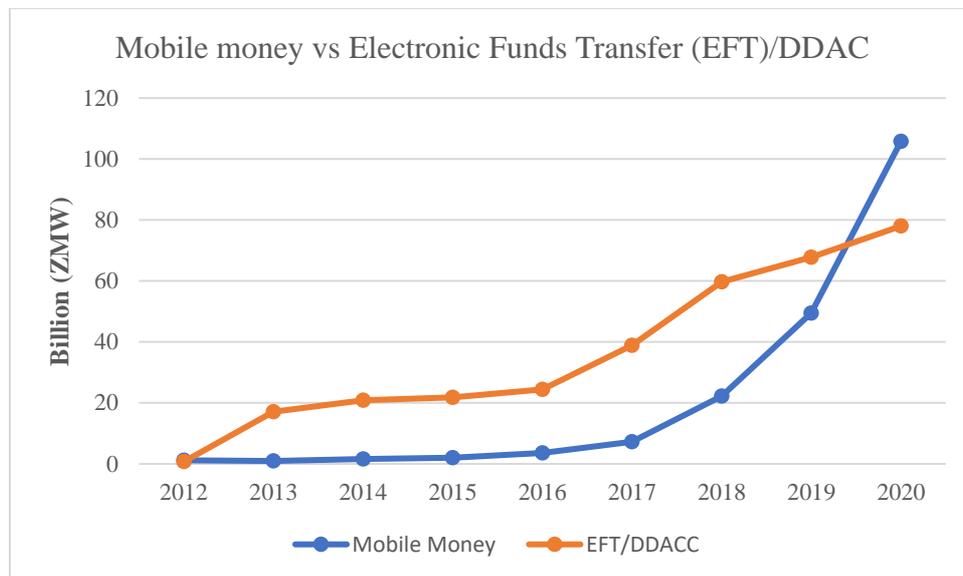

Figure 2. Decade-long (in value terms) Mobile Money vs Electronic Funds Transfer (EFT)/DDAC (BoZ 2021)

This trend has drawn cyberattackers to the mobile money landscape who have sought to maliciously tap into this plethora of funds (Phiri 2019). It is in light of the aforesaid that this paper aims to present a high-order analytical approach towards mobile phone-based phishing attacks in Zambia which seek to defraud benign victims. In order to achieve the foregoing, the following research objectives were put into perspective;

- to ascertain the degree of understanding of the different types of mobile-phone-based phishing attacks in Zambia
- to determine the extent and prevalence of mobile phone-based phishing attacks in Zambia
- to ascertain the implications of clicking on phishing links sent in SMS or WhatsApp messages
- to explore countermeasures against phishing attacks in Zambia based on the uncovered attack properties

The rest of this paper is organized as follows; Section 2 explores the related literature while Section 3 presents the attack model and the applied framework. The adopted methodology is presented in Section 4 whilst the findings and results are presented in Section 5. The discussions thereof come forth in Section 6 whereas the conclusion, together with the recommendations are drawn in Section 7.

## 2. Literature Review

The Zambian ICT sector and cyberspace have been growing rapidly in the past two decades since the liberalisation of the telecoms sector and the introduction of multiple Internet gateways (Zimba et. al, 2020). Unlike the Mwembeshi Earth Station gateway that had limited capacity (Nyirenda and Cropf 2012), the new gateways offer higher capacities which has resulted in an increased online presence, both for corporate and home users. As such, these users have become targets of cyberattacks. However, there is inadequate literature regarding cyberattacks in Zambia. At the time of this writing, and much to our knowledge, there has been no academic research on mobile phone-based cyberattacks in the Zambian ICT sector. This trend can be attributed to the reluctance of governments in the SADC region to prioritise cybersecurity in their national budgets and national agendas (Renaud 2018).

One of the earliest notable cybersecurity incidents reports in Zambia that drew nationwide attention was the defacing of a government website that saw the portrait of the head of state replaced with a cartoon (BBC 1999). However, the case against the perpetrator failed as the suspect could not be convicted because there were no laws in Zambia to address cybercrimes at the time. Since then, the government has sought to introduce cybercrime legislation (BBC 2004), the final product of which is "The Cyber Security and Cyber Crimes Act" of 2021 (GRZ 2021).

Elnaiem (Elnaiem 2019) explores the role of trust and gender in the adoption of mobile money in Zambia using the Technology Acceptance Model. Despite the scarcity of mobile money-related research, the author notes that cybercrime is a challenge in this domain. Particularly, the author points out impersonation attacks targeted towards the Zoona mobile money platform where the attackers would impersonate staff in an effort to solicit for PINs and access to the user's money. Equally, the strides made by ZICTA, Airtel, and MTN to combat such types of cybercrimes are noted. However, the most common types of mobile phone-based cyberattacks such as phishing, SMishing, Vishing are not addressed.

Bruijn et. al (de Bruijn, Butter, and Fall 2017) present an ethnographic study of the complex interrelation between users and non-users of mobile money. Like Elnaiem (Elnaiem 2019), they also note that that trust is a crucial and overarching theme in the mobile money landscape in Zambia which attackers have since time and again exploited. They particularly point out cybercrimes where the attackers cunningly obtain mobile money access codes and stress on the frustrations that most users in developing countries face when trying to recover their money. They note that in most cases, the victims did not recover their money. The study, nonetheless, does not provide the detailed semantics and the types of attacks encountered in mobile money transactions.

Laura (Frederick 2014) looks at the impact of mobile money usage on microenterprise evidence from Zambia and presents evidence of positive net marginal benefits for microenterprises using mobile money in the range between 36% and 74% increase in profits. This could explain why some microenterprises have joined the bandwagon of adopting mobile money payments which has, in turn, turned them into targets of mobile phone-based social engineering attacks.

Phiri and Banda (Norina Phiri and Eliya Banda 2019) investigate mobile money usage patterns across different mobile network operators and note in their study that the most prevalent mobile money service was airtime recharges followed by fund transfers. They noted that mobile money accounts were least used for purchasing and savings. However, especially with the advent of Covid19, the status has changed as mobile money services are used for almost any type of transaction. This paradigm shift to transactions, as earlier stated, has drawn curious cyberattackers to this domain.

Kanobe and Bwalya (Kanobe and Bwalya 2021) sought to lessen the snags associated with mobile money services in developing economies. Their study found several challenges encountered with mobile money not limited to inadequate monitoring of the mobile money agents, insufficient confidentiality, and privacy in financial transactions. Likewise, the use of generic guidelines and policies, third-party involvement in sensitive mobile money activities, and weak staff recruitment policies were cited as having contributed to the challenges. Like other works mentioned above, the details of mobile phone-based cyberattacks are given little attention.

Mwila (Mwila 2020) assesses the cyberattacks preparedness strategy for both the public and private sector in Zambia. His study finds out that these two sectors have low compliance levels with understaffed cybersecurity experts whose roles are mostly assumed by IT professionals. His study could explain why the country is still facing challenges to address cyberattacks in the mobile money services sector.

Susan (Musunga 2019) presents a critique on the effectiveness of cyber laws in Zambia with regards to cybercrimes. Her research argues that even though the Computer Crimes and Misuse Act no.13 0f 2004 criminalized some cybercrimes, it still did not prohibit other major cybercrimes. She further argues that the then Act imposed lighter sentences for offences that require hefty punishments. She additionally argued that the statistics of cybercrime in the country did not reflect the actual severity owing to the fact that most cybercrime-related cases are not reported. This can be said of mobile money-based cybercrimes. It's worth noting, however, that the country has a new Cybersecurity Act that seeks to address the shortfalls in these previous legislation documents.

Table 1 below shows the comparisons with other related works.

Table 1. Comparison with other works

| Attribute/ Work | Quantitative model evaluation | Attack network model | Phishing evaluation framework | Baseline survey | Attacker profile extraction | Phishing attack campaign stats |
|---|---|---|---|---|---|---|
| Elnaiem (2019) | ✓ | ✗ | ✗ | ✓ | ✗ | ✗ |
| Mwila (2020) | ✗ | ✗ | ✗ | ✓ | ✗ | ✗ |
| Obuhuma, & Zivuku (2020) | ✗ | ✗ | ✓ | ✓ | ✗ | ✗ |
| Bruijn et. al (2017) | ✓ | ✗ | ✗ | ✓ | ✗ | ✗ |
| Boateng & Amanor (2014) | ✓ | ✗ | ✗ | ✓ | ✗ | ✗ |
| Proposed Work | ✓ | ✓ | ✓ | ✓ | ✓ | ✓ |

## 3. Attack Model and Evaluation Framework

In this section, we present an attack model representative of mobile phone-based cyberattacks. Furthermore, we formulate an evaluation framework for the chosen attack. Later, we validate our model with a real-world phishing attack that lurked on several WhatsApp group platforms in Zambia by applying the framework to the proposed attack model. SMishing and Vishing attacks are evaluated using data obtained from various respondents of the survey.

### 3.1 The Attack Model

We model the attack process using the philosophy of discrete conceptual units which collaborate to actualize an attack (Arce and Richarte 2003). The components of the attack process are; the *Threat*

*Actor*, *Actions*, *Assets,* and *Goals*. The **Threat Actor** in our model is the social engineering attacker who is capable of using psychology to exploit the victims. He doesn't have to be technically skilled. He performs certain *Actions* in order to acquire *Assets*. **Actions** include selecting a victim Mobile Network Operator (MNO), selecting the appropriate phone number, and sending a phishing message to the victim. **Assets** are the things that the attacker achieves after performing a certain action. Such things include the acquisition of a proxy phone number, a response from a victim like a mobile money PIN code, etc. **Goals** are the final assets that the attacker seeks to achieve. In our exposition, these include the money itself, mobile money PINs or passwords, credit card information, or any other confidential information. The diagram in Figure 3 depicts the formulated attack model.

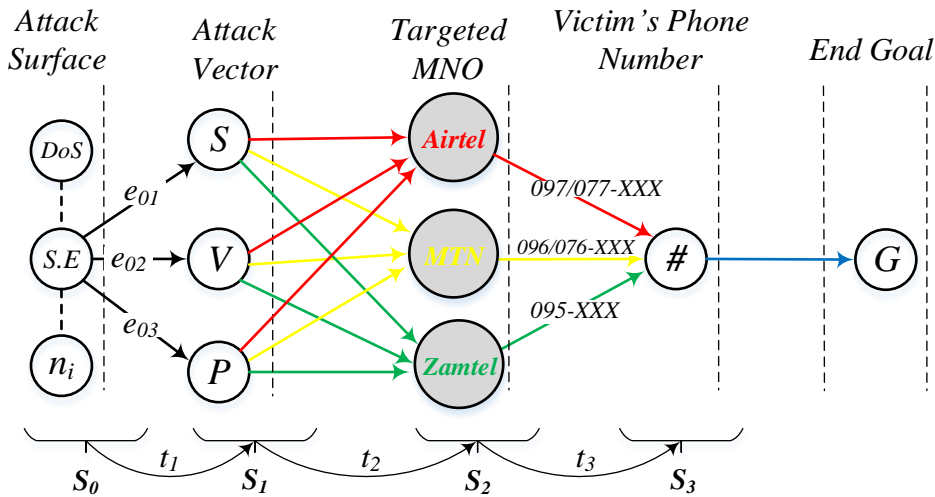

Figure 3. Mobile phone-based social engineering attack model.

The model shows five stages of the attack process namely; Attack Surface, Attack Vector, Targeted MNO, Victim's Phone Number, and the End Goal. The attacker seeks to acquire specific assets in each of these stages. In our exposition, Social Engineering (S.E) is selected from the surface ($S_0$) and the attacker has an option choosing from the following attack vector set: {SMishing (S), Vishing (V), Phishing (P)} denoted in $S_1$. Thereafter, the attacker has to choose which network to find a victim in. In the Zambian scenario, the attacker can choose Airtel, MTN, or Zamtel depicted as $S_2$. After choosing the target MNO, the attacker has the task of finding an operational phone number AND it should be registered for mobile money if the End Goal (G) is to acquire mobile money. The other requirement here is that the phone number should be associated with a smartphone capable of browsing the Internet if the attack vector is phishing through social media platforms such as WhatsApp. This is denoted by # in $S_3$. After that, it is a game of chance that is dependent on the following actions: how convincing the attacker is and how the potential victim responds.

### 3.2 The Phishing Evaluation Framework

Having described the attack process through the proposed attack model, we now turn to formulate a framework for evaluating a chosen attack vector. We devise an evaluation framework for the phishing attack vector because by and large, SMishing and Vishing are considered to be some form of phishing (Yeboah-Boateng and Amanor 2014). The five-step phishing attack evaluation framework is shown in Figure 4.

A typical phishing message usually contains an embedded link. In step 1, we select a phishing message for evaluation. Phishing messages usually implement security-via-obscurity by shortening the embedded link as a Short URL or Tiny URL. In step 2, we extract the Tiny/Short URL embedded in the phishing message to obtain the full URL. In step 3, we extract the full domain name from the complete URL and determine all the domain essentials such as owner of the domain, country where the domain is registered, company where the domain is hosted, etc. Still, in step 3, we subject the URL to reputation engines such as VirusTotal to establish whether such a link has been flagged before by reputable security vendors. In step 4, we implement two types of analyses; dynamic and static analysis/reverse engineering. In dynamic analysis, we run the URL in a sandbox by actually clicking on it to observe its characteristics.

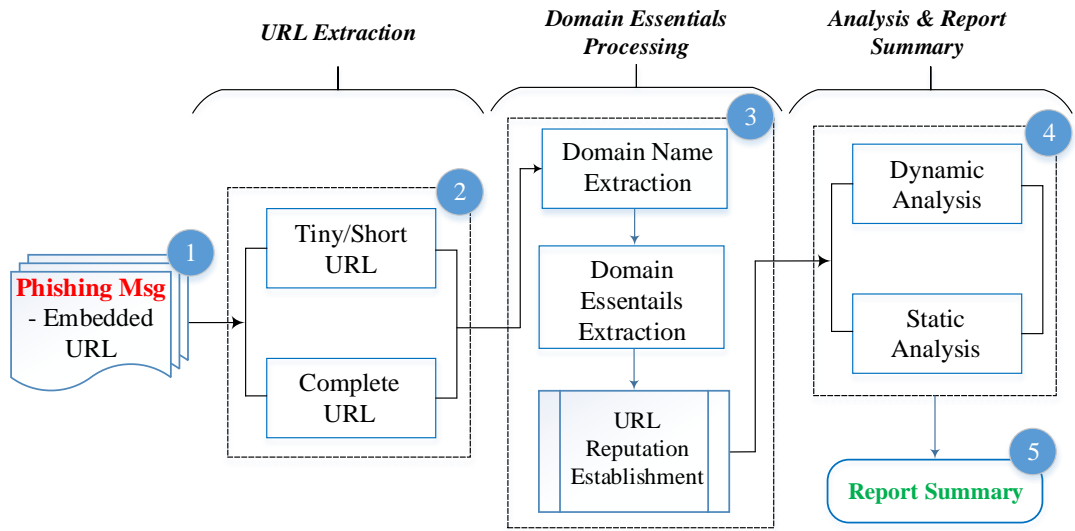

Figure 4. Phishing attack evaluation framework

The REMnux OS was used for sandboxing. In static analysis, we examined the code of the site where the URL redirects. For this purpose, Kali Linux and REMnux were used. Finally, a report summary of the analyses is generated in step 5.

## 4. Methodology

In this section, we present the research methodology of the study which essentially outlines the sequence through which the study was carried out: research design, target population, data collection procedures, and validity/reliability of the research instruments.

### *4.1 Research design*

The research design adopted in this study is a hybrid descriptive research design (Siedlecki 2020) where we use a wide variety of research methods and tools to investigate a variety of variables. As such, unlike in experimental research, we did not seek to manipulate or control any variables but only to observe, discover and measure occurrences. Such an approach provides for an in-depth and deeper understanding of the different facets and dimensions of mobile phone-based cyberattacks in Zambia. Both qualitative and quantitative methods were used together with other detailed technical and analytical methods to explore these attacks in greater detail.

### *4.2 Target population*

Users of mobile phones include a wide spectrum of people, both at corporate and individual levels, young and old alike regardless of gender. As such, the target population for this study included but not limited to personal users and individuals in the public and private organisations, all susceptible to mobile phone cyberattacks. This research used random sampling to select 180 participants. A sample size of 180 with the recommended minimum confidence level of 95% for standard surveys (Feroze and Vinay. Kulkarni 2019) yields a margin of error of 7.25% for a population size of not more than 20,000. Since our target population is above 20,000, the resultant margin of error for the aforesaid confidence level is 7.3% which is within the 4 – 8% acceptable range for standard research (Petersen and Allman 2019).

*4.3 Data collection procedures*

Both qualitative and quantitative data were collected as primary data from the sample through an online survey and face-to-face interviews. The data was then analysed using SPSS and Excel data analysis tools. Charts and other descriptive narratives obtained from these tools were used to outline and highlight the results from which recommendations were drawn.

*4.4 Validity/reliability of research instruments*

To avoid biases, minimise outliers, validate and ascertain the relevance and appropriateness of the survey question to the research objectives, consultations with domain experts were sought. This sought to improve the quality of the research. To ensure the reliability of the survey instruments, the instruments were iteratively piloted to a sample group and the feedback was used to improve the overall structure.

**5. Findings and Results**

In this section, we present two sets of results; evaluation of mobile phone-based phishing attacks based on the model and framework in Section 3.1 and Section 3, and the results of the survey based on the methodology presented in Section 4.

*5.1 Mobile phone-based phishing results*

We followed the steps of the framework in Figure 4. We analysed a phishing message that made rounds in various Zambian WhatsApp groups in early 2021 around April purporting to be a grant from the Zambian government and promising first-time applicants a payment within 24 hours. The message disguises the grant as a relief fund of K10,000. The actual phishing message is shown in Figure 5. This is step 1 in the evaluation framework.

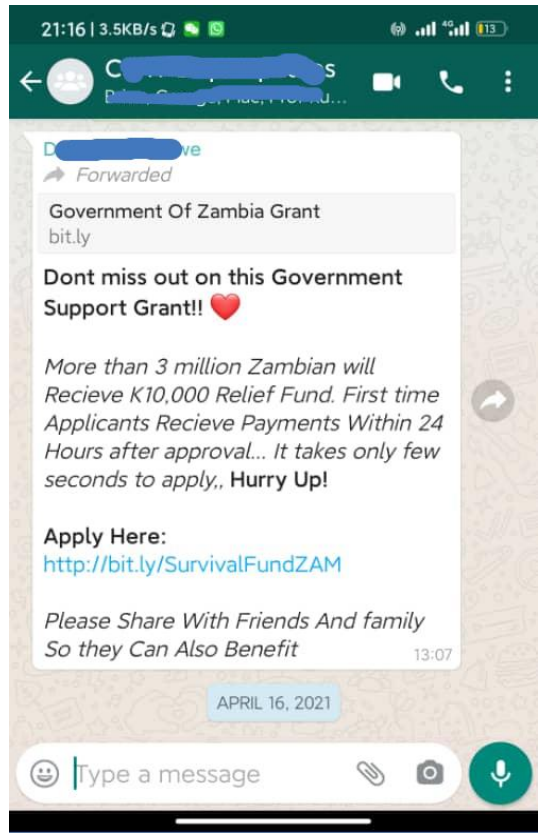

Figure 5. Mobile phone-based phishing message

In step 2, we extract the shortened URL, which in this case is http://bit.ly/SurvivalFundZAM where the victim is coerced to click to apply. Furthermore, we extracted the complete URL for the shortened URL and it resolves to survivalfundzm.blogspot.com. It is worth noting that the initial URL uses plain HTTP whilst the full URL uses HTTPS. Furthermore, the complete URL is not a standalone domain name but rather a subdomain of a BlogSpot site. The exhibit in Figure 6 shows the complete URL and the landing page when the victim clicks on the bait link hoping to apply in order to receive the funds.

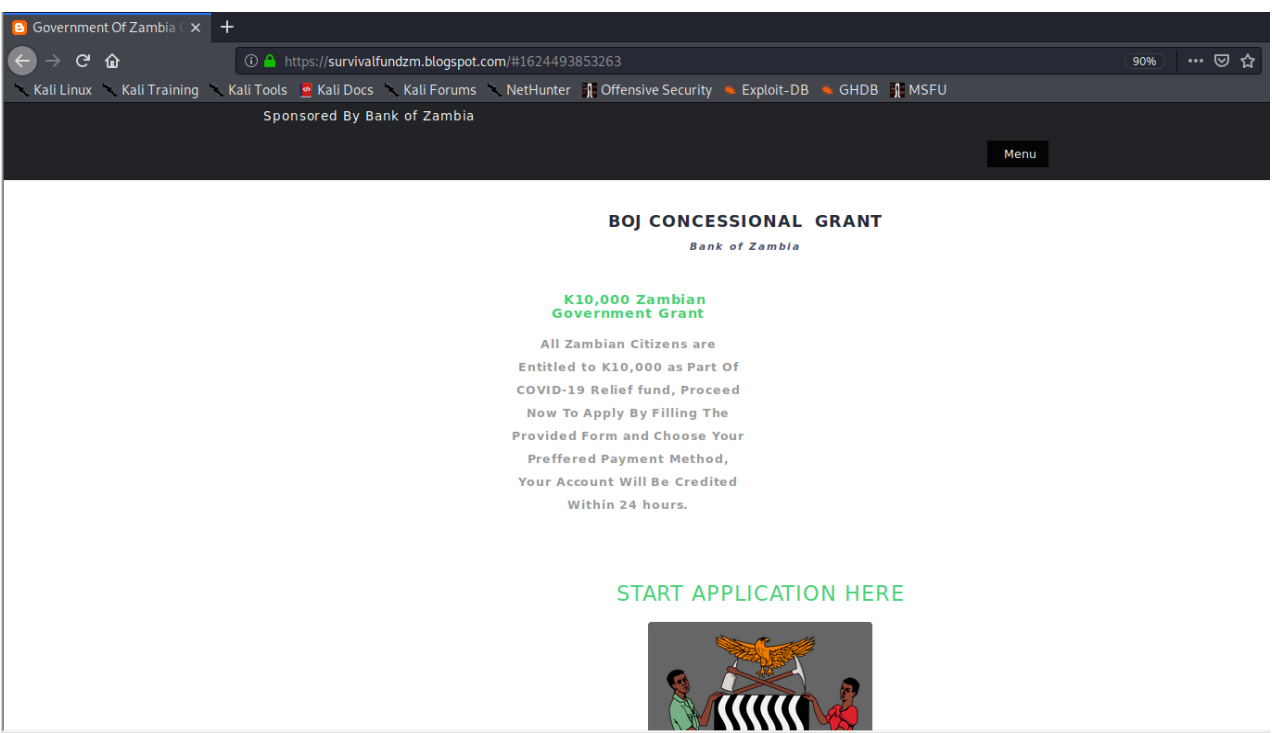

Figure 6. The actual landing page of the phishing link and the complete URL (2021). [Screen Capture]. Retrieved from https://www.survivalfundzm.blogspot.com

Another feature worth mentioning is the use of the Bank of Zambia as the sponsor of the grant and the use of the Zambian coat of arms.

From the extracted domain essentials in step 3, it was discovered that the attacker ran several other phishing scams in other countries. Since Blogger and Google are committed to user privacy and won't avail private information without a police warrant, we instead proceed to check the reputation of the URL on VirusTotal as the final process in step 3. The results are shown in Figure 7.

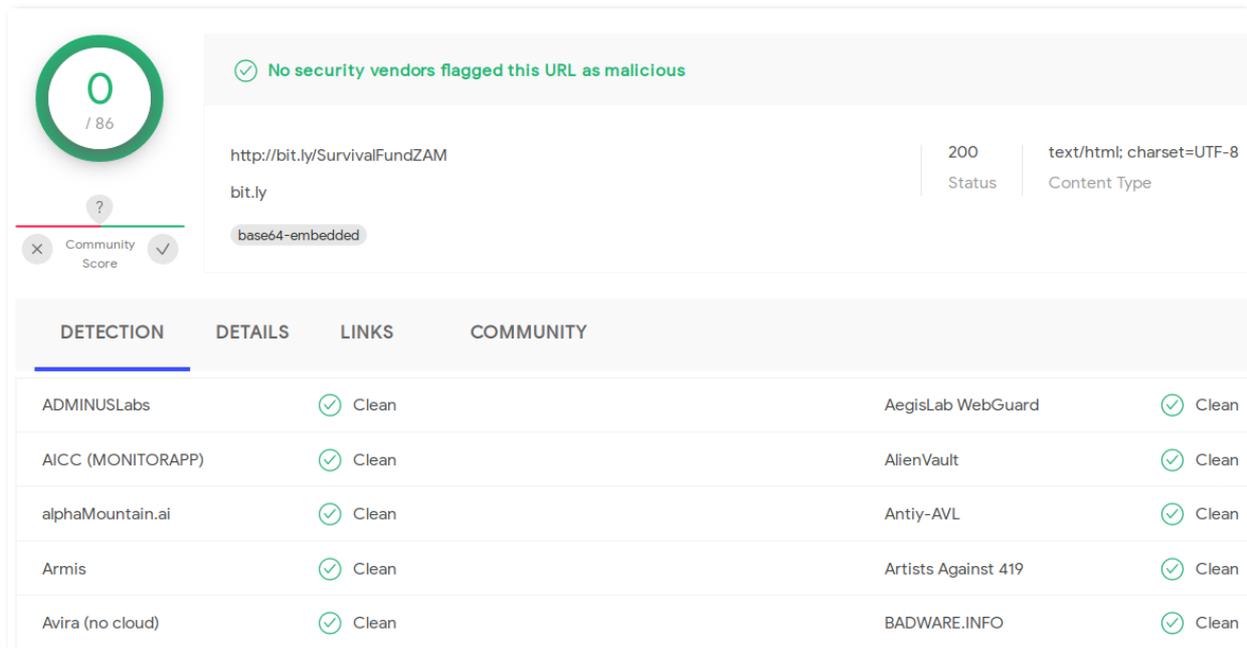

Figure 7. Reputation ranking retrieved from VirusTotal.com. (2021). [Screen Capture]. Retrieved from https://www.virustotal.com/gui/url

Ironically, none of the 86 security vendors flagged the URL as malicious. We provide reasoning for this in the next section. The blog profile of the attacker was extracted and it was discovered that it was created in March 2021.

In step 4, we first start with dynamic analysis where we follow where the link leads whilst running a minimal sandbox. It's in this stage that we encounter the landing page and other pages the URL redirects to. There was no local traffic observed as we avoided filling in the form presented on the landing page. However, we followed the links in the source code paying particular attention to those that capture and process the input data and the lead to another site https://bozgrantfund.blogspot.com. The new redirected site is a fake blog with fake users testifying how they received the money. Furthermore, it gives an impression to the victim that there are already 173,330 users on the site. Using data statistics counters, we obtain web traffic details of web visitors from Zambia to the phishing site as shown in Table 2.

Table 2. Summary statistics of the new redirected link

| Overview | | | |
|---|---|---|---|
| Counting Since: | March 7 2021 | Total | 219,381 Visits |
| Average | 1,903 Visits Per Day | Highest Day | 12,783 / March 13 2021 |
| Today | 465 Visits | Yesterday | 596 Visits |

| This week | 1,693 Visits | Last week | 8,882 Visits |
| This month | 23,302 Visits | Last month | 34,003 Visits |
| This year | 219,381 Visits | Last year | 0 Visits |

| Top Browsers | | |
|---|---|---|
| 1. | Chrome Mobile | 175,089 |
| 2. | Opera Mobile | 22,879 |
| 3. | Mobile Safari | 7,692 |
| 4. | Chrome | 4,396 |
| 5. | Android | 2,631 |

| Top Operating Systems | | | |
|---|---|---|---|
| 1. | Mac OS X | 1 | 5.26 % |
| 2. | Windows 2000 | 1 | 5.26 % |
| 3. | Nokia Series | 1 | 5.26 % |
| 4. | Windows Vista | 1 | 5.26 % |
| 5. | Windows 8 | 1 | 5.26 % |

It's worth noting that the site has only been operational for about three months but has attracted hundreds of thousands of visitors. Even at the time of writing this research, the phishing site was still receiving visitors in the margins of 500 visitors. Further analysis of the details of the visitors and probable victims is shown in Table 3.

Based on these analytics, the number of countries detected to have fallen for the bait is 93. The remainder of the output has been deprecated for presentation purposes. It's alarming to note that Zambia had the highest number of visitors to the phishing site representing about 87.55% of the total visitors. It is expected to have large numbers from Zambia since the phishing message (from Figure 5) harboring these malicious links was targeted at the Zambian audience, hence the use of the Bank of Zambia name and the Zambian Coat of Arms.

Table 3. Distribution of the site visitors' location by country

| Countries | | | |
|---|---|---|---|
| No | | Country or Region | Number of Visitors |
| 1. | | Zambia | 192,069 |
| 2. | | European Union | 21,027 |
| 3. | | South Africa | 1,733 |
| 4. | | United States | 1,010 |
| 5. | | Switzerland | 555 |
| 6. | ss | Nigeria | 452 |
| 7. | | Congo, The Democratic Republic | 345 |
| 8. | | Namibia | 295 |
| 9. | | United Kingdom | 279 |
| 10. | | India | 258 |
| 11. | | Tanzania | 133 |
| 12. | | Kenya | 108 |
| 13. | | Mauritius | 82 |
| 14. | | Zimbabwe | 74 |
| 15. | | Ghana | 73 |

| 16. | Botswana | 62 |
| 17. | Pakistan | 58 |
| 18. | Rwanda | 51 |

After further analyses, we were able to extract the profile of the attacker which revealed that the attacker ran other phishing campaigns targeted at other countries other than Zambia; Jamaica, Ghana, and South Africa. This is depicted in Figure 8.

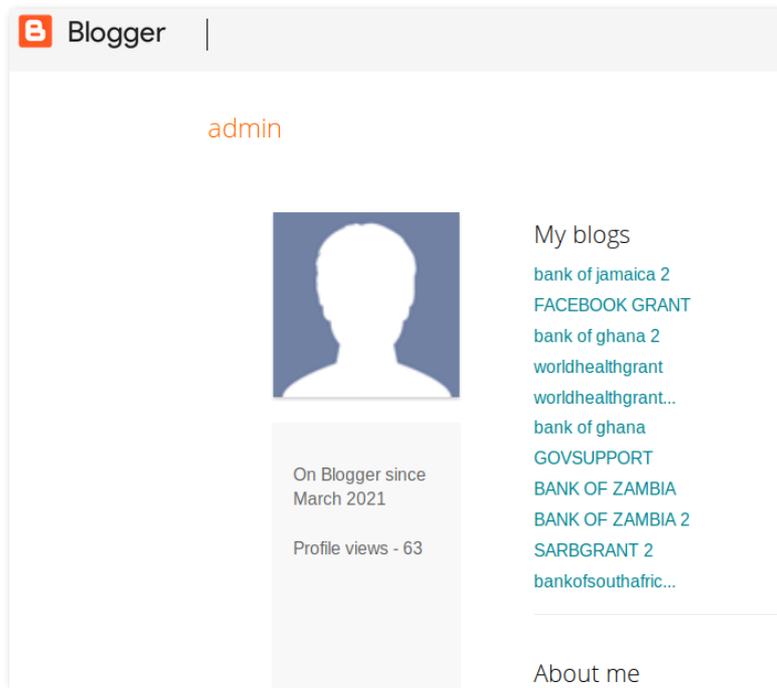

Figure 8. Attacker profile listing other phishing campaigns generated from http://bit.ly/SurvivalFundZAM (2021). [Screen capture]

Additionally, the attacker ran another phishing campaign dubbed Facebook Grant. We finish the static and dynamic analyses of step 4 and proceed to step 5 where we compile a summary report of the evaluation. Details and discussions of this step are expounded in Section 6.

*5.1 Survey results and findings*

Since Phishing, SMishing, and Vishing are attack vectors in social engineering that affect most of the aspects of social media, we sought to assess familiarity with the terminologies themselves. The results are presented in Figure 9 and Figure 10.

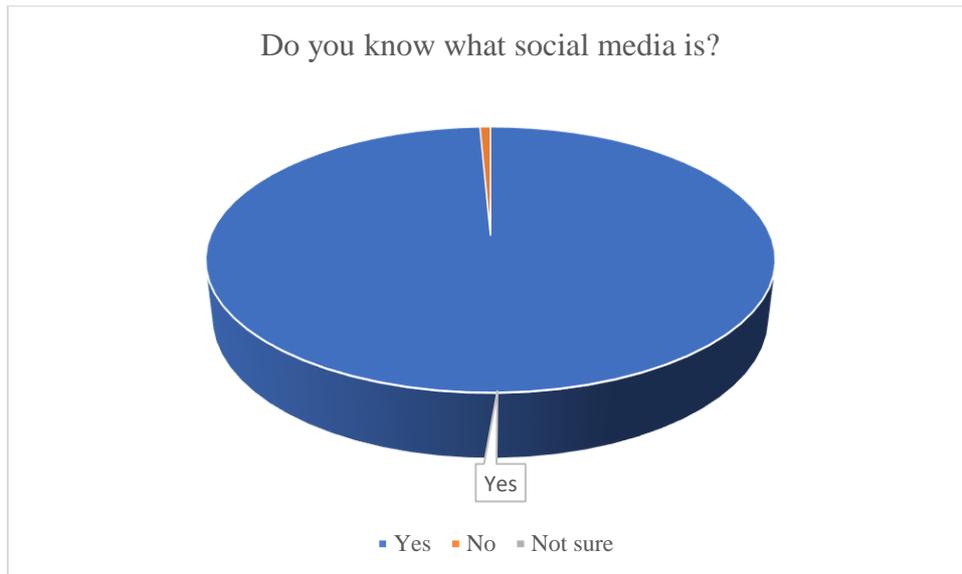

Figure 9. Knowledge of social media

The results in Figure 11 indicate that 99% of the respondents have knowledge of social media. This means that majority of the respondents usually acquire unstructured, mainly tacit knowledge intensively produced, by means of social interactions by users, through social media platforms. Only 1% did not know what it is.

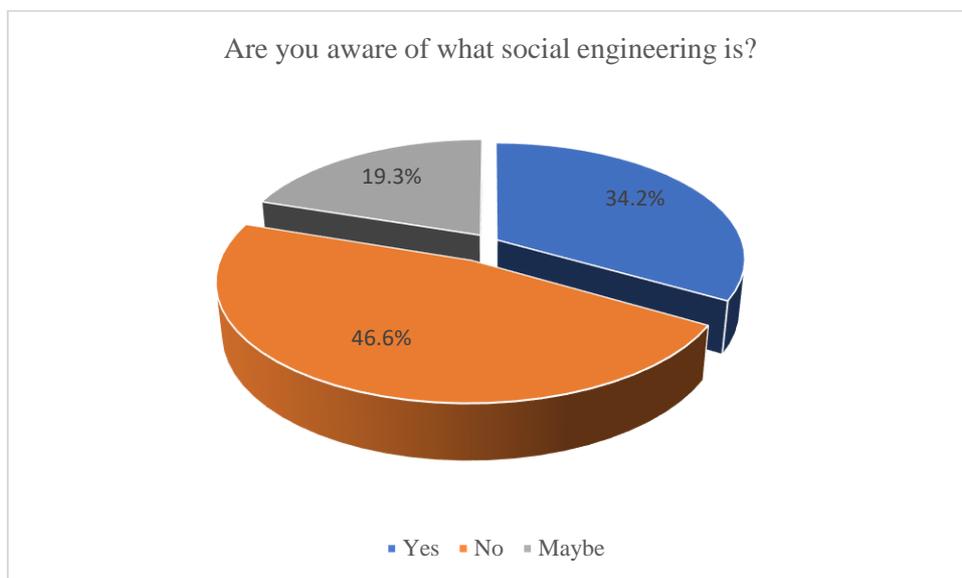

Figure 10. Social Engineering Awareness

In terms of social engineering awareness, only 34.2% of the respondents were aware of what social engineering is. The bigger portion translating to 46.6% of respondents did not know what social engineering is and the remaining 19.3% were indecisive.

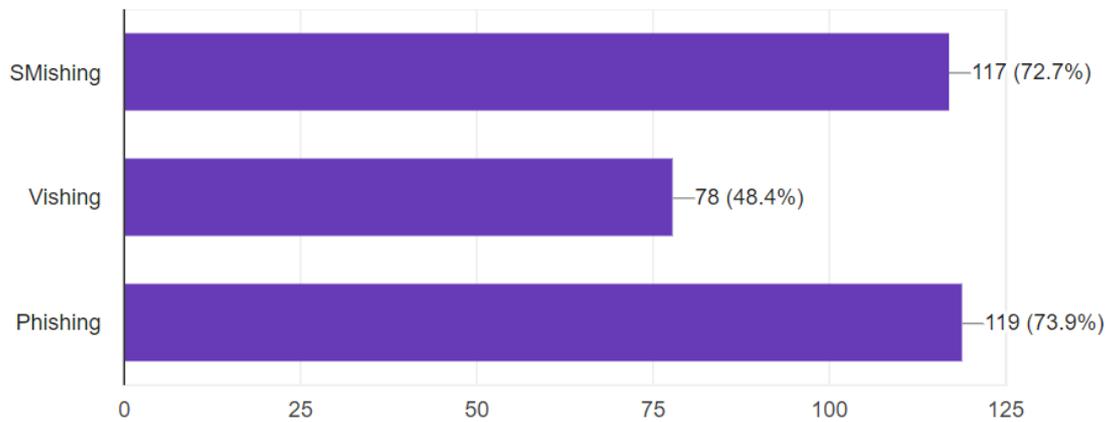

Figure 11. Most Common Fraud Technique amongst SMishing, Vishing or Phishing

When asked which Fraud technique the respondents were familiar with, Figure 11 indicates that Phishing is a Fraud technique that most respondents were aware of representing 73.9% followed by SMishing 72.7% and the least being Vishing representing 48.4%. SMishing is a fraud technique where a person tries to steal money from another through their phone by sending them a fake SMS. Vishing on the other hand is a technique where someone calls the victim pretending to be from a bank or call centre and tries to trick them into sending some money or revealing private information. The Phishing technique applies to a situation where a message with a link is sent to the victim or via WhatsApp group with the promise of winning or getting something when the victim click on it.

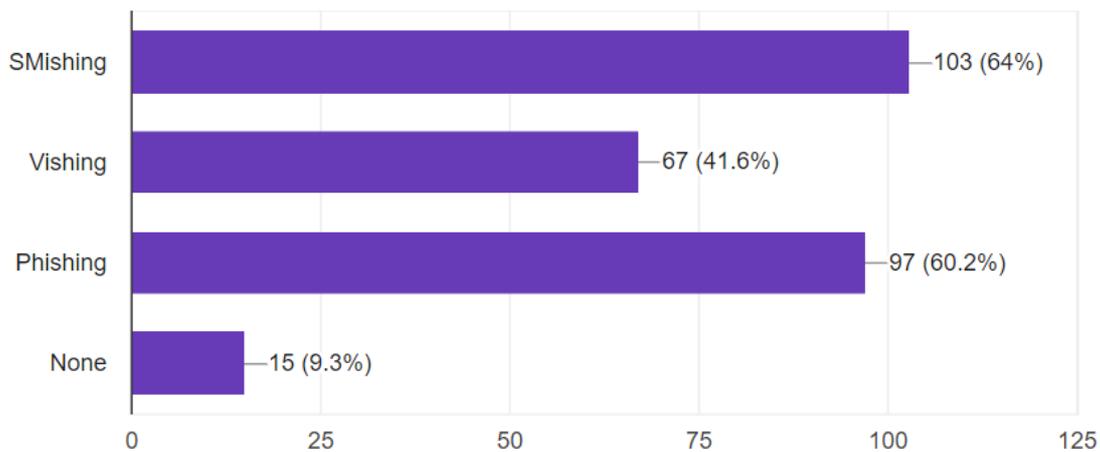

Figure 12. The most commonly encountered type of attack

Most of the respondents were victims of SMishing representing (64%) followed by Phishing (60.2%) and the least being Vishing (41.6%). This means that fake SMS are common when it comes to fraud and there is therefore need to register all the phone numbers by network providers to make it easier to trace the source of the SMS. We can also see in Figure 12 that it was common for victims to receive a message with a link via WhatsApp group or whichever platform promising the victim of winning or getting something good should the victim click on the link.

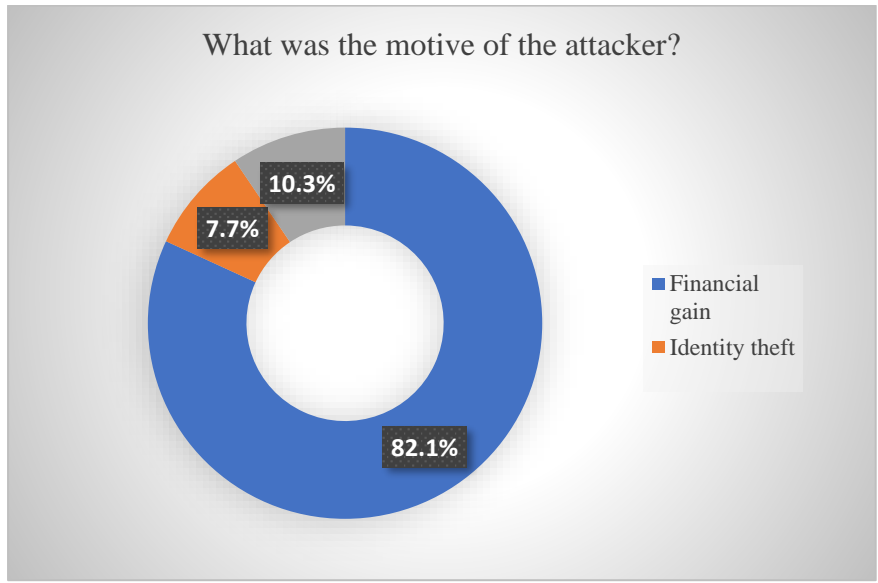

Figure 13. Motive of the attacker

From Figure 13, it can be seen that the motive of the attacker was mainly financial gain translating to 82.1%. Identity theft was in the least with only 7.7% while other motives stood at 10.3%. This explains why the responses in Figure 12 where inclined to SMishing as a commonly used fraud technique where a person tries to steal money from the victim's phone by sending a carefully crafted fake SMS for their financial gain. The target therefore is not the identity of the victim but their finances.

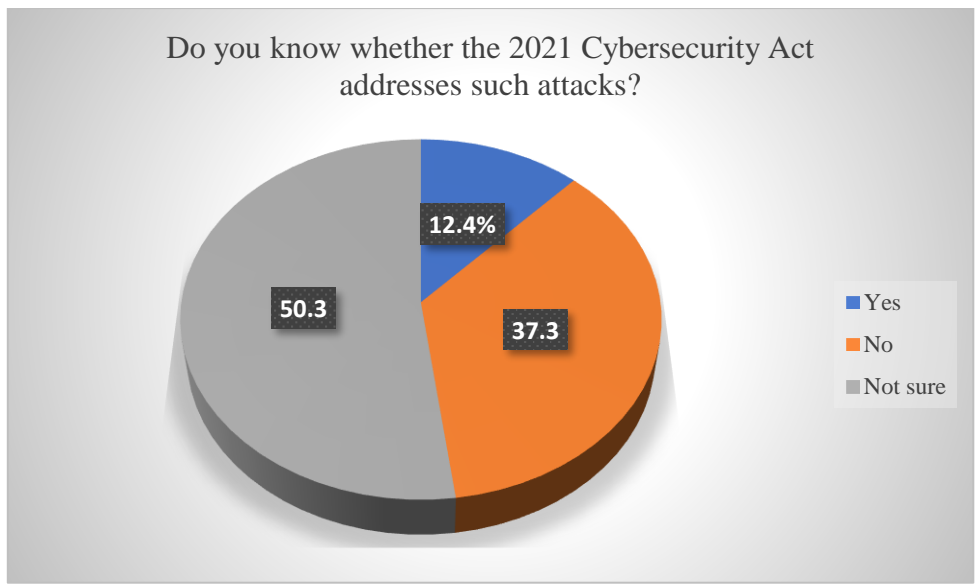

Figure 14. Knowledge on Cybersecurity Act of 2021

We can see in Figure 14 that more than half of the respondents were not sure whether the Cybersecurity Act addresses such attacks. Only 12.4% know that the Act covers social engineering cyberattacks whilst the remaining 37.3% did not know. There is therefore need to enhance this Act so as to achieve its purpose.

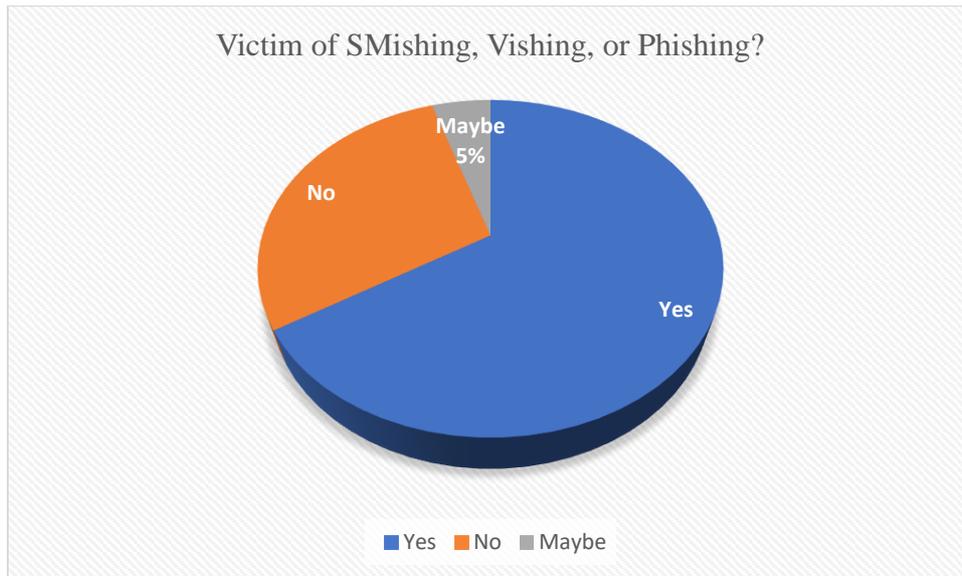

Figure 15. Victims of social engineering cyberattacks

According to Figure 15, we can see that 66% of the respondents have suffered loss (financial or other losses) or know of someone who has suffered loss due to SMishing, Vishing, or Phishing. In fact, 64.6% confirmed having been victims and having incurred some form of a loss.

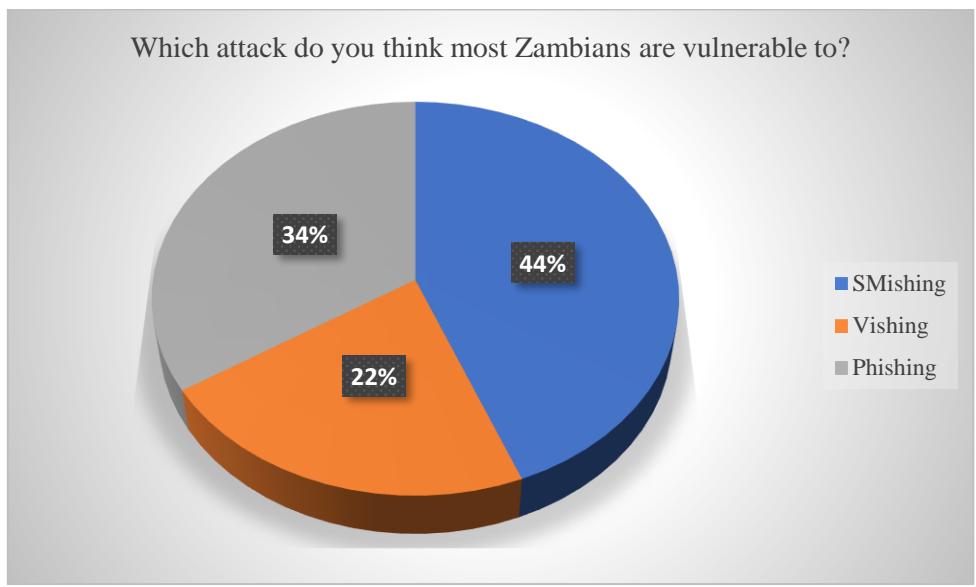

Figure 16. Susceptibility of Zambian mobile phone users to social engineering attacks

All in all, susceptibility of Zambian mobile phone users to social engineering attacks can highly be linked to SMishing at 44% followed by Phishing and Vishing at 34% and 22% respectively. It's worth noting that all mobile phones are capable of voice and text whereas not all of them would be able to process a URL.

### 5.2 Binary Logistic Regression Analysis

To determine the probability of one failing victim to fraud schemes, we perform a binary logistic regression. This is because the response variable is dichotomous that is, it is either you are a victim or not. The model is then;

$$ln\left[\frac{\pi(x_i)}{1-\pi(x_i)}\right] = \beta_0 + \beta_1 x_1 + \beta_2 x_2 + \cdots + \beta_k x_k. \qquad (1)$$

Where
- $\pi(x_i)$= the probability of falling victim to fraud.
- $1 - \pi(x_i)$= the probability of not falling victim to fraud.
- $X_i$= the independent variables in this case being SMishing, Phishing and Vishing.
- $\beta$= the parameter estimates.

Logistic regression is used to predict a categorical (usually dichotomous) variable from a set of predictor variables. With a categorical dependent variable, discriminant function analysis is usually employed if all of the predictors are continuous and nicely distributed; logit analysis is usually employed if all of the predictors are categorical; and logistic regression is often chosen if the predictor variables are a mix of continuous and categorical variables and/or if they are not nicely distributed (logistic regression makes no assumptions about the distributions of the predictor variables). Logistic regression has been especially popular with medical research in which the dependent variable is whether or not a patient has a disease. It is for the same reason this model has been used in this paper because the dependent variable is dichotomous that is; one is either a victim of fraud or not.

**Assumptions:**
- Logistic regression does not assume a linear relationship between the dependent and independent variables.
- The dependent variable must be a dichotomy (2 categories).
- The independent variables need not be interval, nor normally distributed, nor linearly related, nor of equal variance within each group. Probit assume normality in the error terms.
- The categories (groups) must be mutually exclusive and exhaustive; a case can only be in one group and every case must be a member of one of the groups.
- Larger samples are needed than for linear regression because maximum likelihood coefficients are large sample estimates.

One of the many basic concepts in this model is the concept of the ODDS RATIO (OR). An odds ratio is a measure of association between an exposure and an outcome. The OR represents the odds that an outcome will occur given a particular exposure, compared to the odds of the outcome occurring in the absence of that exposure. When a logistic regression is calculated, the regression coefficient for example ß1 is the estimated increase in the log odds of the outcome per unit increase in the value of the exposure. In other words, the exponential function of the regression coefficient $e^{\beta_1}$ is the odds ratio associated with one-unit increase in the exposure. The odds ratio can also be used to determine whether a particular exposure is a risk factor for a particular outcome, and to compare the magnitude of various risk factors for that outcome.

OR is interpreted in the following ways;
- OR=1 Exposure does not affect odds of outcome.
- OR>1 Exposure associated with higher odds of outcome.
- OR<1 Exposure associated with lower odds of outcome.

The odds ratio (OR) is written as follows;

$$\text{Odds=o=}\left(\frac{\pi_i}{1-\pi_i}\right) \qquad (2)$$

- Where $\pi_i$ = the probability of Success.
- $1 - \pi_i$ = the probability of failure.

Regardless of the importance of the odds ratio, our main interest is deriving the probability $P_i$ of an event occurring and this is done as follows;

We defined L = ln(odds of event Y), sometimes called the "log odds" or logit of Y. We can write L in terms of p, Probability, as follows:

$$L = \ln(o) = \ln\left(\frac{P_i}{1-P_i}\right) \tag{3}$$

We can then use the laws of exponents and logs and some algebra to express p (the proportion of successes or risk of the event) in terms of L:

$$\ln(o) = \ln\left(\frac{P_i}{1-P_i}\right) \tag{4}$$

We then exponent both sides,

$$e^L = O = \frac{P_i}{1-P} \tag{5}$$

Multiplying both sides by $(1 - P)$ we get,

$$P_i = e^L(1 - P_i) \tag{6}$$

Making $P_i$ the subject of the formula gives us our end result which is,

$$P_i = \frac{e^L}{1+e^L} \tag{7}$$

The function can be shown in relation to a multiple logistic regression as follows;

$$P_i = \frac{e^{\beta_0+\beta_1 x_1+\beta_2 x_2+\cdots+\beta_k x_k}}{1+e^{\beta_0+\beta_1 x_1+\beta_2 x_2+\cdots+\beta_k x_k}} \tag{8}$$

This is called the logistic function and its graph is as follows. Notice that p, the probability of the event increases from 0 to 1.

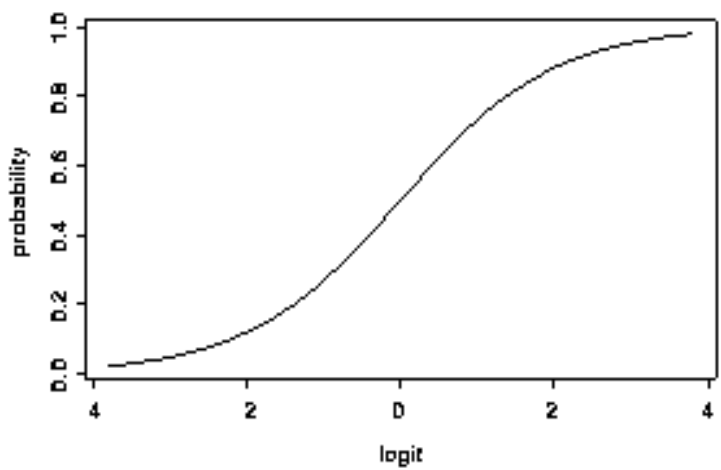

Figure 17. Logit Function

*5.3 Binary Logistic Regression Analysis*

The data consisted of three predictor variables or independent variables, which are SMishing, Phishing and Vishing. The dependent variable was "Are you a victim of Fraud?" which was dichotomous with two responses only i.e. either you are a victim or not.

Table 4. Case Processing Summary

| Unweighted Cases[a] | | N | Percent |
|---|---|---|---|
| Selected Cases | Included in Analysis | 142 | 100.0 |
| | Missing Cases | 0 | .0 |
| | Total | 142 | 100.0 |
| Unselected Cases | | 0 | .0 |
| Total | | 142 | 100.0 |

a. If weight is in effect, see classification table for the total number of cases.

Table 5. Classification Table[a]

| | | | Predicted | | |
|---|---|---|---|---|---|
| | | | Are you a fraud victim | | Percentage Correct |
| Observed | | | yes | no | |
| Step 1 | Are you a fraud victim | yes | 94 | 48 | 66 |
| | | no | 48 | 6 | 34 |
| | Overall Percentage | | | | 80.0 |

a. The cut value is .500

The classification table suggests that if we knew nothing about our variables (i.e. SMishing, Phishing and Vishing) and guessed that a person would be a victim to fraud, we would be correct 80% of the time.

Table 6. Variables not in the Equation

| | | | Score | df | Sig. |
|---|---|---|---|---|---|
| Step 0 | Variables | SMishing | 1.569 | 1 | .210 |
| | | Phishing | 5.094 | 1 | .024 |
| | | Vishing | 9.455 | 1 | .002 |
| | Overall Statistics | | 13.550 | 3 | .004 |

The variables not in the equation tell us whether each independent variable improves the model. The answer is yes for all the variables, as all are significant and if included would add to the predictive power of the model. If they had not been significant and able to contribute to the prediction, then termination of the analysis would obviously occur at this point.

We then conduct the omnibus tests of model. It can be seen from the table below that the model is significant with a p-value of 0.000

Table 7. Omnibus tests of Model Coefficients

|        |       | Chi-square | df | Sig. |
|--------|-------|------------|----|------|
| Step 1 | Step  | 25.261     | 3  | .000 |
|        | Block | 25.261     | 3  | .000 |
|        | Model | 25.261     | 3  | .000 |

The Model summary is conducted and is vital to the interpretation of this regression. Although there is no close analogous statistic in logistic regression to the coefficient of determination, $R^2$, the Model Summary below provides some approximations. *Cox and Snell's R-Square* attempts to imitate multiple R-Square based on 'likelihood', but its maximum can be (and usually is) less than 1.0, making it difficult to interpret. The table is shown below;

Table 8. Model Summary

| Step | -2 Log likelihood | Cox & Snell R Square | Nagelkerke R Square |
|------|-------------------|----------------------|---------------------|
| 1    | 42.373[a]         | .271                 | .475                |

a. Estimation terminated at iteration number 20 because maximum iterations has been reached. Final solution cannot be found.

Here it is indicating that the logistic model explains 27.1% of the variation in the DV. The *Nagelkerke* modification that does range from 0 to 1 is a more reliable measure of the relationship. *Nagelkerke $R^2$* will normally be higher than the *Cox and Snell* measure. *Nagelkerke $R^2$* is part of SPSS output in the 'Model Summary' table and is the most reported of the R-squared estimates. In our case it is 0.80, indicating a moderately strong relationship of 80% between the predictors and the prediction.

An alternative to model chi square is the **Hosmer and Lemeshow** test, which divides subjects into 10 ordered groups of subjects and then compares the number actually in the each group (observed) to the number predicted by the logistic regression model (predicted). The 10 ordered groups are created based on their estimated probability; those with estimated probability below 0.1 form one group, and so on up to those with probability 0.9 to 1.0. Each of these categories is further divided into two groups based on the actual observed outcome variable (success, failure). The expected frequencies for each of the cells are obtained from the model. The table is shown below;

Table 9. Contingency Table for Hosmer and Lemeshow Test

|        |   | doyouhavereadymarket = yes | | doyouhavereadymarket = no | | Total |
|--------|---|----------|----------|----------|----------|-------|
|        |   | Observed | Expected | Observed | Expected |       |
| Step 1 | 1 | 3        | 6.885    | 8        | 5.115    | 12    |
|        | 2 | 3        | 2.731    | 4        | 5.269    | 8     |
|        | 3 | 3        | 1.681    | 8        | 10.319   | 12    |
|        | 4 | 0        | .266     | 8        | 7.734    | 8     |
|        | 5 | 0        | .285     | 12       | 11.715   | 12    |
|        | 6 | 0        | .135     | 8        | 7.865    | 8     |
|        | 7 | 0        | .017     | 12       | 11.983   | 12    |
|        | 8 | 0        | .000     | 8        | 8.000    | 8     |

A probability (p) value is computed from the chi-square distribution with 8 degrees of freedom to test the fit of the logistic model. If the H-L goodness-of-fit test statistic is greater than 0.05, as we want for well-fitting models, we fail to reject the null hypothesis that there is no difference between observed and model-predicted values, implying that the model's estimates fit the data at an acceptable level. That is, well-fitting models show non-significance on the H-L goodness-of-fit test. This desirable outcome of non-significance indicates that the model prediction does not significantly differ from the observed. The output in table is shown below;

Table 10. Hosmer and Lemeshow Test

| Step | Chi-square | df | Sig. |
|---|---|---|---|
| 1 | 8.174 | 6 | .226 |

The H-L statist*ic* assumes sampling adequacy, with a rule of thumb being enough cases so that 95% of cells (typically, 10 decile groups times 2 outcome categories = 20 cells) have an expected frequency > 5. Our H-L statistic has a significance of 0.226 which means that it is not statistically significant and therefore our model is quite a good fit.

The variable in the equation table is the most important of all tables. It includes the coefficients of the independent variables and their odds ratios. The Variables in the Equation table has several important elements. The Wald statistic and associated probabilities provide an index of the significance of each predictor in the equation. The Wald statistic has a chi-square distribution.
The simplest way to assess Wald is to take the significance values and if less than .05 reject the null hypothesis as the variable does make a significant contribution. In this case, we note that all the variables contributed significantly to this model. The variables in the equation table is shown below;

Table 11. Variables in the Equation

|   |   | B | S.E. | Wald | df | Sig. | Exp(B) |
|---|---|---|---|---|---|---|---|
| Step 1[a] | SMishing | 18.158 | 11563.752 | .000 | 1 | .999 | 76868443.811 |
|   | Phishing | 10.776 | .584 | 9.239 | 1 | .002 | 5.906 |
|   | Vishing | 2.124 | .669 | 10.087 | 1 | .001 | .120 |
|   | Constant | 13.398 | 11563.752 | .000 | 1 | .999 | .000 |

a. Variable(s) entered on step 1: SMishing, Phishing, Vishing.

Table 12. Variables in the Equation

|   |   | B | S.E. | Wald | df | Sig. | Exp(B) |
|---|---|---|---|---|---|---|---|
| Step 0 | Constant | 1.735 | .313 | 30.690 | 1 | .000 | 5.667 |

The Exp(B) column in the table presents the extent to which raising the corresponding measure by one unit influences the odds ratio. We can interpret EXP(B) in terms of the change in odds. If the value exceeds 1 then the odds of an outcome occurring increase; if the figure is less than 1, any increase in the predictor leads to a drop in the odds of the outcome occurring. For example, the EXP(B) value associated with SMishing is 76868443.811. Hence, with one unit rise in the odds ratio, SMishing is 76868443.811 times as large and therefore one is 76868443.811 times likely to be schemed by SMishing. Therefore, SMishing **is a key factor.**

The 'B' values are the logistic coefficients that can be used to create a predictive equation. It is fitted in the same manner a linear regression model is fitted. Our model can be seen below as follows:

**Formula: Standard Logit Model**

$$ln\left[\frac{\pi(x_i)}{1-\pi(x_i)}\right] = \beta_0 + \beta_1 x_1 + \beta_2 x_2 + \cdots + \beta_k x_k. \quad (9)$$

This is the standard logit model. After placing the coefficients, the model now becomes;
**Formula: Fitted Model**

$$ln\left[\frac{\pi(x_i)}{1-\pi(x_i)}\right] = 13.398 + 18.158 SMishing + 10.776\ Phishing + 2.124\ Vishing \quad (10)$$

From the above fitted model, it can be seen that all factors have a positive relationship with one failing victim to fraud with $SMishing$ being more dominant.

## 6. Discussions of Results

A lot of insights were gained from the results presented in the previous section. In this section, we discuss those insights starting with phishing and then ending with SMishing and Vishing.

### *6.1 Phishing*

Evaluation of the phishing attack based on the presented model and framework revealed that a lot of Zambians fell victim to the attack campaign. Although we cannot tell without further analysis what details the attacker got away with, we can for sure tell that the Zambian mobile users visited the phishing site than others. With the phishing campaign for Zambia starting on 7th March 2021, Zambia stood at 87.55% with a total of 192,307 out of 219,664 visits. The highest number of visits recorded per day was 12,783 on 13th March 2021. This implies that most Zambians fell for the scam within a week of its initialisation. South Africa on the other had a total of 2,147 out of 2,939 thus representing a 73.05%. Meanwhile, for the same type of attack campaign in Ghana, victims who clicked on the link accounted for 121,957 out of 103,906 representing an 85.21% victim rate. Despite Zambia still leading generally in terms of being highly victimised, it is important to note that Ghana and South Africa have higher populations than Zambia. If we compare the 192,307 Zambian victims against the country's population and do the same with South Africa with only 2,147 victims but yet with a way higher population than Zambia, we understand that the attack statistics are actually more severe in Zambia. The same trend is observed if Zambia is compared with Ghana on those realistic criteria.

The top browsers used to visit the phishing site by users from Zambia were Chrome Mobile at 79.81%, Opera Mobile at 10.43%, and Mobile Safari at 10.43%. This shows that the victims must have received the phishing message on their phones and visited the malicious site using their phones since all these browsers are mobile phone browsers. Chrome and Opera Mobile browsers are commonly used in Android-based phones such as Samsung, Huawei, Infinix, iTel, Redmi, etc while Safari Mobile is usually used in Apple devices such as iPhone. This indeed is reflective of the distribution of the types of phones common in Zambia.

The susceptibility of Zambian mobile phone users could be attributed to many things. Nonetheless, bringing users up to speed with regard to cybersecurity is not an activity that can be done in isolation. Awareness campaigns and user training ought not to be bureaucratic and isolated because the weakest link in the attack chain, the human being, is affected by a lot of factors that will determine their approach to cybersecurity. As such, apart from educational background and level of technical knowledge, social and cultural factors ought to be put into consideration when formulating measures to address these cybersecurity threats.

*6.2 SMishing and Vishing*

The results from the survey show that more than 99% of mobile phone users understand the term *social media* but only 34% of them know what social engineering attacks are. This hints at the level of awareness of the population. Most of the respondents were aware of the different tactics that attackers use in Phishing, SMishing, and Vishing and the data shows that most of the respondents have been victims of SMishing despite being more aware of phishing. Vishing was not common amongst the respondents. This could in part be attributed to the fact that Vishing involves actually calling the victim and would more likely be successful if the attacker has some background information. This possibility has led to some mobile money agents being accused of conspiring with the attackers (Malakata 2019). In either case, the attacker has to use an operational phone number and since all phone numbers have to be registered in Zambia, attackers are alleged to use inaccurately and incorrectly registered SIM cards. The diagrams in Figure 18 (a) – (d) show typical SMishing messages from operational phone numbers across different networks. We remove any identifying information such as phone numbers and names with which those numbers have been registered for privacy purposes. However, we leave the MNO to show that SMishing attacks are prevalent on all networks. We only retain the first three digits to assist in determining the mobile network that is being used. This is echoed in the attack model in Figure 3.

In Figure 18(a), the attacker is using the Airtel network and sending the SMishing message to an Airtel phone number. The attacker banks on the possibility that the potential victim is about to send Airtel money to someone. If the victim is less suspecting or a little careless for one reason or the other, they will end up sending the money to the attacker as was in one case (Phiri 2019). In Figure 18(b), the attack logic is identical to that in Figure 18(a) only that the attack this time is taking place on the MTN network.

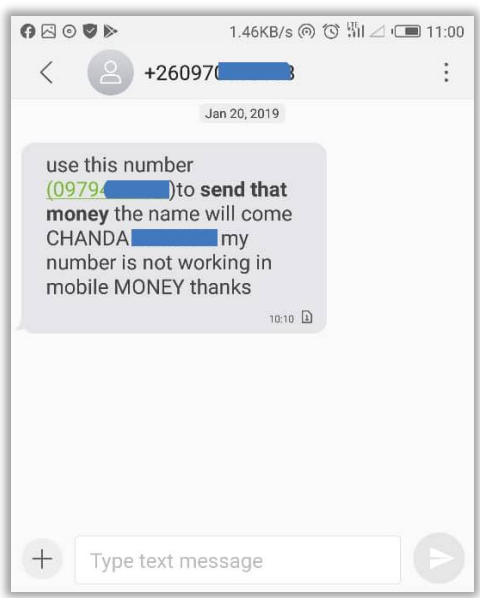 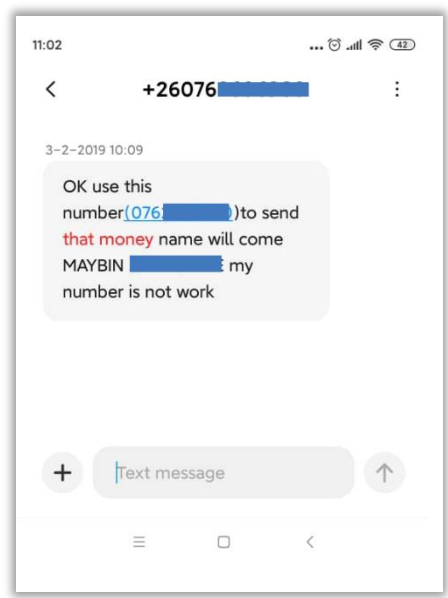

Figure 18 (a). SMishing on the Airtel network    Figure 18 (b). SMishing on the MTN network

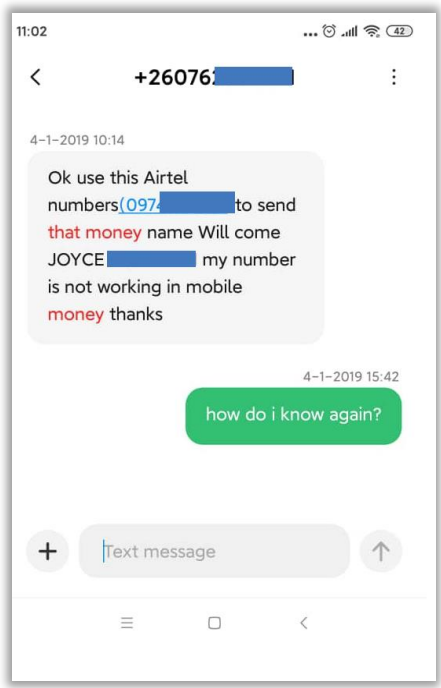 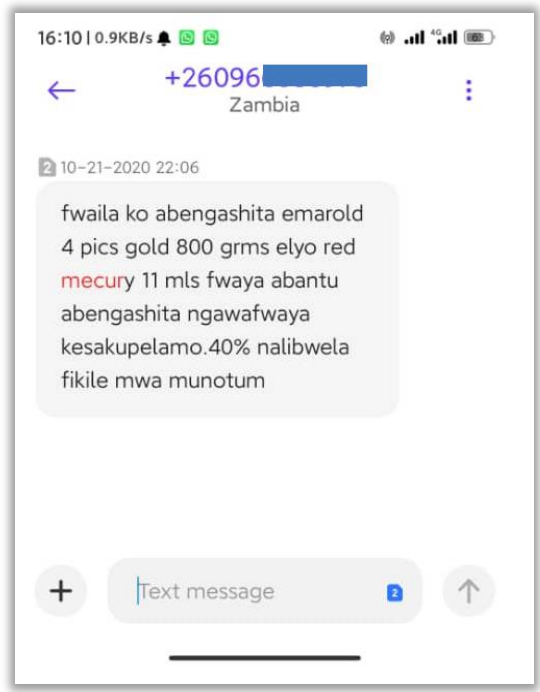

Figure 18 (c). SMishing across MTN and Airtel networks

Figure 18 (d). SMishing in a local language on the MTN network

The attack in Figure 18(c) adds another dimension in that the attacker is on the MTN network but is coercing the victim to send money to a number on the Airtel network. There are many possibilities why the attacker would choose a cross-network attack approach but since that is beyond the scope of this work, we leave it to future works.

On the other hand, the attack in Figure 18(d) presents a different SMishing attack paradigm in two ways. First, the attacker is not asking for mobile money directly but is seeking to engage the potential victim and exploit them further depending on how vulnerable they are. Second, the attacker is using a message crafted in a local language (Bemba). In this attack, the attacker is alleging to have four pieces of gold weighing 800g and 11ml of red mercury. He further promises to give 40% of the sales to the potential victims which in itself sounds too good to be true. Unfortunately, a lot of citizens have fallen for such schemes.

Vishing on the other hand was not so common from the survey. However, those that had been targeted mentioned that the attackers spoke good English and tried to authority in the conversation. In most of the cases, the attackers were after private information such as PIN codes and passwords. They would impersonate officials mostly from banks and MNO. Some of them would even send some instructions via SMS after the call. Most of the attackers used ordinary phone numbers and others are said to have had some foreign accent. When probed further using a local language, the attack would insist on speaking English or eventually cutting the call.

Furthermore, it was observed that among the ways used to convince victims was false statements that MTN/Airtel/Zamtel has a reward for the victim and that they need their mobile money password for one to get the reward. Others informed the victim that they had won money or other items but need bank details like passwords to receive the money. At times, the attacker would pretend to be a law enforcer and wanted the victim's help to solve a crime. Others were promised employment, sent a fake e-wallet message making the victim believe they have received the money. The schemer would send bank-like messages after which the victim received a phone call to confirm the message. Others were informed that they had won a prize but to collect it, they needed to have a small amount in their account or had to send a small fee and were given instructions making the victim believe they are receiving money when in fact money ended up being withdrawn from their own account to the attacker's.

In both types of attack, the attack spends some considerable resources since SMS and voice calls cost money. Since SMS are generally cheaper than ordinary voice calls, and that coupled with the fact that the attackers won't expose themselves via voice, it could explain why SMishing is more prevalent as opposed to Vishing.

Unlike in phishing where the attacker is usually an outsider to the country, SMishing and Vishing are perpetrated by fraudsters from within the country. This is usually evidenced by the fact that the sender's phone number (in the case of SMishing) and the caller's phone number (in the case of Vishing) have always been from one of the local MNO, i.e., Airtel, MTN, or Zamtel. As such, the mitigation approach towards SMishing and Vishing should not be lumped together with phishing. They all have to be isolated and considered case by case.

## 7. Conclusions and Recommendations

This research examined the level and degree of exposure of Zambian mobile phone users to mobile phone-based cyberattacks that are usually implemented via social engineering. SMishing stands out as the most prevalent form of social engineering attacks amongst mobile phone users closely followed by phishing. Vishing is the least common type of attack and this is attributed to several factors not limited to cost and probability of exposure of the attacker. However, since it's easier to convince someone over the phone as they will not have enough time to reason through during the phone call, Vishing might yield more results even though it's the least common type of attack.

The emergency of these types of attacks is not only attributed to the widespread adoption of mobile phones, the introduction of mobile money but incompleteness of the law, people's low awareness of fraud prevention, and the strength of government regulation among other things. The motive behind the attacks was financial gain although identify theft was common for Vishing attacks. Lack of awareness and user training were cited as some of the leading reasons why victims fall for these types of attacks. Based on the insights from this research, the following are some of the suggested solutions against these types of attacks:

- Awareness and education of the users about the attacks and prevailing laws against them
- Enactment of unambiguous laws
- Thorough enforcement of laws
- Stiffer penalties and blacklisting of perpetrators from MNOs

Another suggestion that came out strong is that the authorities should motivate users by making known incidents where perpetrators were pursued and prosecuted and convicted. This could be achieved via a dedicated unit that periodically gives information on successful prosecution and convictions. Other details to be considered in updating the users on the latest schemes employed by the attackers as the tactics tend to evolve over time.

**Declaration of interests**

The authors declare that they have no known competing financial interests or personal relationships